\newif\ifANONYMOUS \ANONYMOUStrue \ANONYMOUSfalse
\newif\ifCOMMENTS \COMMENTStrue 

\documentclass[11pt,a4paper]{article}

\usepackage{amsfonts,amssymb,amsthm,bm} %
\usepackage[longnamesfirst]{natbib}
\bibpunct{(}{)}{;}{a}{,}{,}
\usepackage[normalem]{ulem} %
\usepackage[tbtags]{amsmath} %
\usepackage{enumerate}
\usepackage{fullpage}
\usepackage[hidelinks]{hyperref}
\usepackage{booktabs, adjustbox}

\usepackage{setspace}
\usepackage[font=small,labelfont=bf]{caption} 
\usepackage{graphicx,color}
\graphicspath{{./slides/_pics/}{./pics/}}

\definecolor{BrickRed}{rgb}{.625,.25,.25}

\definecolor{markergreen}{rgb}{0.6, 1.0, 0}
\definecolor{darkgreen}{rgb}{0, .5, 0}
\definecolor{darkred}{rgb}{.7,0,0}

\theoremstyle{plain}
\newtheorem{theorem}{Theorem}

\theoremstyle{definition} 
\newtheorem{definition}[theorem]{Definition}

\usepackage{mathrsfs}

\newcommand{\E}{{\mathbb{E}}}

\providecommand{\R}{{\mathbb{R}}}

\newcommand{\p}{{\bf P}}

\newcommand{\dd}{{\rm d}}

\providecommand{\Ncdf}{{\rm N}}

\newcommand{\1}{\ensuremath{\mathbf{1}}}

\ifx\prop\undefined
\newtheorem{prop}{Proposition}[section]
\fi

\newcommand{\argmax}{\operatornamewithlimits{argmax}}

\begin{document}
\title{\Large\bf Correlation scenarios and correlation stress
  testing\thanks{\protect\linespread{1}\protect\selectfont  
We acknowledge helpful comments from Martin Aichele, Carol Alexander, Michael
      Eichhorn, Bertrand Maillet, Radu Tunaru, two anonymous referees
      as well as participants at the 2nd Financial Economic Meeting:
      Post-Crisis Challenges 
      2021, the MathFinance Digital Conference 2021, the CQF
      Institute, the FAST Seminar at the University of Sussex, and the
      Research Seminar in Statistics and Mathematics at WU Vienna. }
\thanks{Declarations of interest: none}
    \thanks{\protect\linespread{1}\protect\selectfont
      This research was supported by the Deutsche Forschungsgesellschaft (DFG)
    through the International Research Training Group 1792 ``High
    Dimensional Nonstationary Time Series''.} 
}
\ifANONYMOUS
\author{Anonymous manuscript for peer review}
\else
\author{
N.\ Packham\thanks{Berlin School of Economics and Law,
  \href{mailto:natalie.packham@hwr-berlin.de}{natalie.packham@hwr-berlin.de}}\
\ 
and 
F.\ Woebbeking\thanks{Goethe University Frankfurt,
  \href{mailto:woebbeking@finance.uni-frankfurt.de}{woebbeking@finance.uni-frankfurt.de}}} %
\fi

\maketitle %

\begin{abstract}
\noindent We develop a general approach for stress testing
correlations of financial asset portfolios.
The correlation matrix of asset returns is specified in a
  parametric form, where correlations are represented as a function of
  risk factors, such as country and industry factors. A sparse factor
  structure linking assets and risk factors is built using Bayesian
  variable selection methods.
Regular calibration yields a joint distribution of
economically meaningful stress scenarios of the
factors. As such, the method also lends
itself as a reverse stress testing framework: using the
Mahalanobis distance or highest density regions (HDR) on the
joint risk factor distribution allows to 
infer worst-case correlation scenarios. We give examples of stress
tests on a large portfolio of European and North American stocks.  
\end{abstract}

\medskip 


\ifCOMMENTS


\fi

\clearpage
\section{Introduction}\label{sec:introduction}

\noindent Correlation is one of the most important, if not the
  most important risk factor in finance, driving everything from na\"ive diversification to
the effectiveness of hedges. It 
is well-established that correlations fluctuate
over time and may be strongly affected by specific events
\citep{Karolyi1996,longin2001extreme,Ang2002a, wied2012testing,
  pu2012correlation, adams2017correlations}. Changes in correlation
may lead to potentially unexpected or unquantified losses, see e.g.\
LTCM \citep{Jorion2000}, Amaranth Advisors \citep{Chincarini2007},
JPMorgan's ``London Whale'' \citep{Packham2019}. Regulators have since
called for a better correlation risk management.\footnote{See for
  example the European Capital Requirements Regulation (CRR): Articles
  375(1), 376(3)(b), 377.} However, a unified and generally
  accepted correlation risk management framework does not yet exist.

Building on the work of \citet{Packham2019}, this article develops a
universal framework for building realistic correlation scenarios,
correlation stress testing and reverse stress testing. 
The novelty of this approach is the link between correlations and
observable risk factors. ``Classical'' stress testing determines the
impact on a financial position as a consequence of changes in risk
factors, see e.g.\ \citep{EBA2021}. Here, we consider the impact on
asset correlations of a risk factor scenario.
With this framework, one can challenge
diversification benefits based on economic scenarios. Likewise, one
can assess the effectiveness and risks of hedges based on economic
scenarios. In particular, correlation stress tests
produce extreme, yet plausible correlation scenarios from
economically relevant risk factors. Our 
method therefore addresses: first, the selection of appropriate
correlation risk factors for a given portfolio; second, the
parameterization of large correlation matrices and their mapping into
risk measures; third, the identification of critical scenarios through
reverse stress testing. The method is particularly relevant for
supervisors who  ``\textit{are considering the ways in which stress
  tests can be integrated best into the regulatory framework}''
\citep{pliszka2021system}. Reverse stress testing complements other
stress testing methods, such as stressed
  value-at-risk, introduced in the Basel 2.5 framework
  \citep{BIS2011}. Stressed VaR determines portfolio risk using the
  parameters of a time period considered as stressful. Reverse stress
  testing, on the other hand, determines the risk factor configuration
  that is both stressful and plausible for a given portfolio.

  Contrary to extremely prudent scenarios where diversification effects
  are ignored, we are particularly interested in establishing a link
  between risk factors and assessing the plausibility
  of scenarios. It is this link that enables reverse stress testing,
  i.e., the identification of critical risk factor scenarios, which,
  from the point of view of a practitioner, may see as the most
  valuable aspect of stress testing. 

A widespread method in financial risk management and in asset
  management is to {\em capture\/} correlations through factor models,
  see e.g.\ \citep[Section 3.4]{McNeil2005}. A common choice for the   
  factors are industries and geographic regions. In this article,
  correlations are {\em  modelled\/} by linking each
  asset to a subset out of the set of 
  factors. Asset correlations are specified in a functional form,
  where the correlation between any two assets is determined by the
  shared, resp.\ unshared links. The degree by which shared or
  unshared links affect correlations is obtained by calibrating
  ``weight'' parameters to empirical correlations.  Scenarios are
generated by varying these parameters. 

Given the history of calibrated parameters, the 
method lends itself to reverse stress testing as it is capable of
identifying the factor structure of worst case scenarios. More
specifically, given the mapping of correlation 
risk factors to a risk measure, one can find the global maximum of the
risk measure and infer the corresponding risk factor scenario.  As
each parameter represents an economically relevant correlation risk
factor, it is therefore possible to identify critical portfolio 
structures (``smoking guns'') that might require particular attention
from a risk management perspective. 

Plausibility, or lag thereof, is a common problem for scenarios that
are generated through reverse stress testing. This article addresses
plausibility by assigning a joint probability distribution to the
correlation parameters, which in turn allows to quantify the
plausibility of correlation  
scenarios. In this article, the constraint is specified via the \emph{Mahalanobis distance} and {\em highest
  density regions} (HDR), both of which can be thought of generalising
the concept of a quantile to a multivariate setting, as will be
  explained below.  

Assigning the appropriate correlation risk
factors to an asset is a delicate exercise: ignoring relevant risk factors might lead to undetected
correlation risk, while selecting an excessive amount of risk factors
potentially renders the economic interpretation meaningless, as
  the impact of specific risk drivers becomes indistinguishable across
  assets. A
good factor selection mechanism should therefore focus on a
  sparse selection of relevant factors for each asset. In
addition, since typically 
  some persistent information about the relationships between
  assets and factors is available (such as the country of the
    headquarter and the main industry), the factor selection method
  should offer the ability to incorporate prior knowledge.
{\em Bayesian variable selection methods\/} support both of
  these requirements by allowing to specify prior information as well
  a giving control over the number of factors selected.

  The framework developed here can be implemented for any
    financial application that bears correlation risk, such as asset
    allocation or hedging. As an example, we apply the factor-model approach to a large
and well-diversified equity portfolio. For this particular portfolio,
geographic regions and industries serve as
correlation risk factors.

A further application where correlation scenario and stress testing
can reveal inherent risks is the practice of so-called ``portfolio
margining'' in initial margin calculations of clearing houses. Here,
netting of offsetting positions reduces the margin
requirement. However, if positions hedging each other are not
  perfect substitutes, but only
highly correlated, then significant de-correlation
could lead to
substantial margin calls, thereby increasing counterparty risk at a
systematic level.

The literature on the role of correlation and dependence in
  finance is vast, but interestingly the literature on
establishing correlation 
stress tests is comparably scarce.
  It is well established that correlations are not constant
over time and may be strongly affected by specific events
\citep{longin2001extreme, wied2012testing,
  pu2012correlation}. \citet{adams2017correlations} observe that
correlations vary over time and, in addition, experience level shifts
and structural breaks that occur in response to economic or financial
shocks. The regime switching behaviour of financial asset
correlations, especially in times of crisis, is confirmed and at the
heart of several studies, including \citet{Junior2012, Buccheri2013,
  Papenbrock2015}. \citet{krishnan2009correlation} and
\citet{mueller2017international} provide empirical evidence that
investors demand a correlation risk premium, which is related to the
uncertainty about future correlation
changes. \citet{buraschi2010correlation} develop a framework for
inter-temporal portfolio choice that includes hedging components
against correlation risk. \citet{Tumminello2010, Keskin2011} analyse
the topology of correlation matrices as networks or hierarchical
trees, providing insights on the structure, taxonomy and hierarchy of
financial asset correlations. This last strand of the literature
is probably closest to our framework in the sense that it provides an
interpretation of the factors driving correlation changes.

\citet{Packham2019} introduce a correlation stress testing
  methodology tailored to the case of the so-called {\em ``London
    Whale''\/}, a USD 6.2 billion loss on a credit derivatives
  portfolio at JPMorgan. The loss was partly due to the de-correlation
  of positions that were supposed to act as hedges for each
  other. Correlation stress testing would have revealed this risk
  early on and might therefore have led to a more prudent assessment
  and de-leveraging of the position. The factors in the ``London
  Whale'' case were tailored to match characteristics of the credit
  derivatives, such as their maturity, credit quality (investment
  grade versus high yield) and geographic origin (CDX in the US,
  iTraxx in Europe).

The prominent role of correlation in financial portfolios has
led 
regulatory agencies to call for risk model
stress tests that account for ``significant shifts in correlations''
\citep[p.~207 ff.]{bcbs128}.
However, there is little literature on parametric correlation
modelling, an exception being parametric functions for
  correlation and volatility in interest rate modelling (e.g.\ LIBOR
  market model), see
  \cite{Rebonato2002,Brigo2002,Schoenmakers2003}.
  Another strand of the literature on correlation stress testing
  deals with the question of ensuring positive semi-definiteness of
  the matrix, see \citet{higham2002computing,
  qi2010correlation, ng2014black}. 


The selection of plausible scenarios poses a challenge in the
development of stress testing methods in general. The use of
historical or hypothetical scenarios is problematic, as the
probability and thus the plausibility of a
scenario is unknown and relevant scenarios
might be neglected. In an extensive study, \citet{alexander2008developing}
compare various well-known models in their ability to conduct
meaningful stress tests.  \citet{glasserman2015stress} develop an
empirical likelihood approach for the selection of stress scenarios,
with a focus on reverse stress testing. \citet{kopeliovich2015robust}
present a reverse stress testing method to determine scenarios that
lead to a specified loss level.
\citet{Breuer2009} and \citet{flood2015systematic} use the Mahalanobis
distances to select scenarios from a multivariate distribution of risk
factors. 

\citet{breuer2013systematic} extend these approaches and consider
various application scenarios, amongst them stressed default
correlations, which refer to the correlations of Bernoulli variables
denoting the default or survival of loans or
obligors. \citet{Studer1999} considers correlation breakdowns by
identifying the worst-case correlation scenario in a constrained
region of P\&L scenarios. However, solving the problem turns out to be
computationally intractable.\footnote{More precisely, the optimisation problem is NP-hard, i.e., requires a non-deterministic polynomial computation time.}
  Also, the likelihood or
plausibility of such a correlation scenario is not known. The
difference in our setting is that we model correlation itself in a
parametric way and -- imposing a risk factor distribution calibrated from historical data --
find the risk-factor scenario that produces the worst loss within a
given plausibility region.

In summary, we contribute to the literature by proposing a flexible
  correlation stress testing framework that is capable of adapting to
  different requirements and settings. A central feature of our method
  is the Bayesian selection of correlation risk factors. Choosing the
  ``right'' factors is a timely and relevant exercise, especially from a
  regulatory perspective.\footnote{Banks, for example, present
      capital models that contain factor models to
      regulators. Approving these factor models is challenging as the
      literature provides little guidance on how to check if the
      factors in the model are well chosen.} Furthermore, we explore
  how reverse stress testing can help to construct and understand
  extreme yet plausible risk factor scenarios. This article tests the
  method on a large equity portfolio; however, the method would easily
  adopt to any other financial portfolio.

The article is structured as follows: Section~\ref{sec:corr-stress-test}
lays out the factor model for correlation stress
testing. Section~\ref{sec:fact-select-stock} introduces the
 Bayesian correlation factor selection
mechanism. Section~\ref{sec:stress-test-scen} applies the stress
testing framework to a large stock portfolio; and
Section~\ref{sec:conclusion} concludes.

\section{Correlation stress testing methodology}
\label{sec:corr-stress-test}

This section presents the correlation stress testing
  framework. First, we define a factor model structure on
  correlations, which establishes a correlation matrix under
  stress. Second, we infer the impact of the stressed correlations
  by calculating value-at-risk with the modified correlation
  matrix. Finally, by imposing a probability distribution on the
  factors driving correlation, we determine plausible worst-case
  correlation stress scenarios. The last step is commonly known as
  reverse stress testing.


\subsection{Factor model}\label{sec:factor-model}

An economically meaningful correlation stress testing framework
requires linking correlations with risk factors, such as economic
variables or financial market indicators. In the context of portfolio
allocation or risk management, the factors could represent industries
and countries. While \citet{Packham2019} applied correlation stress
testing in a very specific context, this article aims to develop a
generic and flexible correlation stress testing framework, intended to
work in different contexts and for different applications.
  
Consider a portfolio of $p$ assets and assume that there are $d$ risk
factors. Each asset is associated with a number of these risk
factors. We will introduce details on how factors are assigned in
Section \ref{sec:fact-select-stock}; it
  should be noted, however, that for a stress test to
be meaningful, the number of factors associated with an asset needs to
be sufficiently small.

The association of asset $i\in \{1,\ldots, p\}$ with factor
$k\in \{1,\ldots, d\}$ is denoted by the indicator variable
$\1_{\{k,i\}}$.
The correlation of asset returns $i$ and $j$ is
modelled as
\begin{equation}
  \label{eq:1}
  c_{ij} = \text{tanh}\Big(\eta + \underbrace{\sum_{k=1}^d \lambda_k
    |\1_{\{k,i\}}-\1_{\{k,j\}}|}_{\text{``inter''-correlations}}
  + \underbrace{\sum_{k=1}^d \nu_k \1_{\{k,i\}}
    \1_{\{k,j\}}}_{\text{``intra''-correlations}} \Big), 
\end{equation}


with coefficients
$\eta, \lambda_1, \ldots, \lambda_d, \nu_1, \ldots, \nu_d\in \R$ and
$\text{tanh}:\R\mapsto [-1,1]$ the tangens hyperbolicus. 
Aside from conveniently mapping to $[-1,1]$ and being monotone
increasing, the main motivation for choosing
the function $\text{tanh}$ is its use in inferential statistics on
sample correlation coefficients.\footnote{The argument of the
  $\text{tanh}$ function, $z:=\text{arctanh}(c_{ij})$ is the so-called
  {\em Fisher $z$-transformation} \citep{Fisher1915,Fisher1921}. See
  also e.g.\ \citet{Casella2002} and \citet{Remillard2016}. \citet{Fisher1921}
  shows that if $c_n$ is the sample correlation determined from an
  $n$-sized sample of a bivariate normal distribution with correlation
  $|\rho|<1$, then
  $\sqrt{n} (z_n-\alpha) \stackrel{\mathcal L}{\longrightarrow}
  \Ncdf(0,1)$ as $n\rightarrow\infty$.} The following summation
formula serves as useful a approximation for the interpretation of
individual coefficients, especially if the
coefficients are close to zero:
\begin{equation}
  \tanh(x+y) = \frac{\tanh x + \tanh y}{1+\tanh x  \tanh y} \approx
  \tanh x + \tanh y. \label{eq:tanhsum}
\end{equation}

The constant $\eta$ can be thought of as 
a ``base'' correlation.\footnote{Due to multicollinearity issues, it may be necessary to omit
    the constant.} The
coefficients 
$\lambda_1, \ldots, \lambda_d$ model ``inter-factor'' correlations:
the higher $\lambda_k$, the higher the correlation impact if exactly
one of the two assets is associated with the $k$-th factor (in
  which case $|\1_{\{k,i\}}-\1_{\{k,j\}}|=1$). Similarly,
$\nu_1, \ldots, \nu_d$ express ``intra-factor'' correlation: the
higher $\nu_k$, the higher the correlation between assets
jointly exposed to factor $k$ (since in this case
  $\1_{\{k,i\}} \1_{\{k,j\}}=1$). The concept of ``inter''- and
  ``intra''-correlations is found in the context of credit risk in
  e.g.\ \citep{Duellmann2008}. 



Given a sample correlation matrix at one point in time, the
coefficients
$\eta, \lambda_1, \ldots, \lambda_d, \nu_1, \ldots, \nu_d$ can be
determined e.g.\ by ordinary least squares on
$\text{arctanh}(c_{ij})$, the inverse of $\text{tanh}$.  Simple
correlation scenarios such as ``the correlation between assets exposed
to factor $k$ and assets not exposed to factor $k$ increases'' is then
implemented by increasing $\lambda_k$. Likewise, a scenario such as
``the correlation of firms exposed to factor $k$ increases'' is
implemented by increasing $\nu_k$.  With time series of historical
data, the coefficients can be calibrated on a regular basis, from
which resonable scenarios can be determined.


The matrix defined by Equation~\eqref{eq:1} is not guaranteed to
  be positive semidefinite and thus may need to be further transformed
  to yield a valid correlation matrix.  Converting a
  non-positive-semidefinite matrix
  $\tilde{C} \in \mathbb{R}^{n \times n}$ into a positive-semidefinite
  matrix can be approached as a matrix nearness problem, with
  nearness expressed by a suitable norm, such as the Frobenius
  norm (sum of absolute difference of all matrix
  entries). \citet{higham2002computing} provides an algorithm that
  finds the correlation matrix satisfying
  $\min \{||\tilde{C} - C || : C \text{ is a correlation matrix} \}$
  by exploiting the spectral properties of $\tilde C$.
It may also be possible that a correlation matrix fails to be
  positive semidefinite due to computational precision, in which case
  the eigenvalues of the matrix are just slightly below zero. In this
  case it can be sufficient to transform the matrix as 
  $C = (1-\epsilon) \tilde{C} + \epsilon I$, where $I$ is the identity
  matrix and $\epsilon$ is a small constant.




\subsection{Stress testing}\label{sec:stress-test}

With a portfolio's value being just the sum the constituents'
  values, a ceteris paribus
  shift in correlation has no {\em instantaneous\/} effect on the
value. Therefore, to reveal the impact of a correlation
stress test requires calculating portfolio risk measures. 
In the simplest setting, portfolio risk is measured by value-at-risk
(VaR) in a {\em variance-covariance approach}, i.e., 
\begin{equation}
  \label{eq:var}
  \text{VaR}_\alpha = -\Ncdf_{1-\alpha} \cdot  V_{0} \cdot
  \left(\mathbf w^\intercal\, \boldsymbol\Sigma\, \mathbf w\right)^{1/2},
\end{equation}
where $\Ncdf_{1-\alpha}$ denotes the $(1-\alpha)$-quantile of the
standard normal distribution, $V_{0}$ denotes the current position
value, $\mathbf w$ is the vector of portfolio weights and
$\boldsymbol\Sigma$ denotes the covariance matrix of the portfolio
returns. In this setting we
assume that the expected return is zero, which is a common and
  reasonable
assumption for short time horizons. 

The normal distribution assumption can easily be generalised, e.g.\ to
a Student $t$-distri\-bution. The $\text{$t$-VaR}$ is
discussed in \citet{Packham2019}, together with the possibility to
jointly apply correlation and volatility stress scenarios. In fact, 
any model or method that takes a correlation matrix as an input is
suitable for the correlation stress testing approach.
%

\subsection{Reverse stress testing}\label{sec:reverse-stress-test}

When stress testing, aside from understanding the impact of given
scenarios, one is also interested in the opposite question: What is
the worst scenario amongst all scenarios that occur within some
pre-given range? In a univariate setting, one would select a quantile
of the risk factor distribution -- this is the principal idea
underlying value-at-risk. Different extensions of quantiles to a
multivariate setting exist, for example the {\em Mahalanobis
  distance\/} \citep{Mahalanobis1936}, {\em highest density regions
  (HDR)\/} \citep{Hyndman1996} or concepts based on norms, see e.g.\
\citep{Serfling2002}.

The {\em Mahalanobis distance\/} of a vector $\mathbf X\in \R^n$ with
expectation $\mathbf \mu\in \R^n$ and covariance
$\boldsymbol\Sigma\in \R^n\times\R^nn$ is defined as
$D(\mathbf x,\mathbf \mu, \boldsymbol\Sigma) = (\mathbf x-\mathbf
\mu)^\intercal \boldsymbol\Sigma^{-1} (\mathbf x-\mathbf\mu)$, see
e.g.\ \citep{Mahalanobis1936,Kent1979,McNeil2005}.  The Mahalanobis
distance is an appropriate measure for reverse stress testing if the
underlying distribution is elliptic or at least symmetric. Moreover, if
$\mathbf X$ is normally distributed, then
$D^2(\mathbf X)\sim \chi^2(n)$, where $n$ is the length of
$\mathbf X$, which greatly simplifies identifying if scenarios lie
within a range specified by a probability.

The highest-density region (HDR) \citep{Hyndman1996} generalises the
idea of the Mahalanobis distance to arbitrary probability
distributions. It is defined as the region with probability $1-\alpha$
that has the smallest possible volume in the sample space;
equivalently, every point inside the region should have probability
density at least as large as every point outside of the
region. Formally, if $f(\mathbf x)$ is the density of $\mathbf X$,
then the $(1-q)$--HDR is the subset $R(f_q)$ of the sample space such
that $R(f_q)=\{x: f(x)\geq f_q\}$, where $f_q$ is the largest constant
such that $\p(\mathbf X\in R(f_q))\geq 1-q$.

A straightforward way to calculate the HDR is via Monte Carlo
simulation, see \cite{Hyndman1996}. Note that
$\p(f(\mathbf X)\geq f_q)=1-q$; in other words, $f_q$ is the
$q$-quantile of $f(\mathbf X)$. Given an iid sample of observations,
$f_q$ can be estimated from the sample, and all samples in the HDR are
easily identified by having a density value greater than $f_q$. If the
sample size is large enough, then a reduction of the estimator's
variance is achieved by a variant of Latin Hypercube Sampling (LHS)
targeted at dependent random vectors \citep{Packham2010}.

In order to allow for skewness and more variation in tail heaviness
than a normal distribution, we calibrate the time series of
coefficients
$\boldsymbol \beta = (\eta, \lambda_1,\ldots, \lambda_d, \eta_1,
\ldots, \eta_d)^\intercal$ to a multivariate normal inverse Gaussian
(NIG) distribution (see Appendix \ref{sec:norm-inverse-gauss})%
\footnote{The NIG distribution can be thought of a
    generalisation of the multivariate normal distribution, allowing
    for skewness and more variation in the tail while still being
    light-tailed, which appears appropriate for the parameters.}  and
infer reverse stress test scenarios as the worst scenarios within the
HDR at a given level $\alpha$, i.e.,
\begin{equation}
  \boldsymbol\beta^\ast=\argmax_{\{\boldsymbol\beta \in R(f_q)\}}
  \text{VaR}_\alpha(\boldsymbol\beta),  \label{eq:betastar}
\end{equation}
where $\text{VaR}_\alpha$ is given by Equation (\ref{eq:var}) with
correlation matrix imposed by $\boldsymbol\beta$.  From
Equation~(\ref{eq:var}), it is obvious that maximising
$\text{VaR}_\alpha$ does not depend on $\alpha$ and is equivalent to
maximising the variance. A trivial consequence is that
$\boldsymbol\beta^\ast$ also maximises expected shortfall
$\displaystyle \text{ES}_\alpha = \frac{1}{1-\alpha} \int_\alpha^1
\text{VaR}_u\, \dd u$.


  \section{Factor selection}
  \label{sec:fact-select-stock}

%

As mentioned in
  the introduction, \citet{Packham2019} demonstrate the correlation stress testing
  methodology tailored to the case of the so-called {\em ``London
    Whale''\/}, where the factors were manually chosen to match characteristics of the underlying credit
  derivatives portfolio, such as their maturity, credit quality (investment
  grade versus high yield) and geographic origin (CDX in the US,
  iTraxx in Europe).

  In general, the choice of factors will depend on the type of
  correlation stress test to be conducted, and the assignment of
  relevant factors to assets may not be as straightforward as in the
  ``London Whale'' case. In a standard credit risk or asset
  allocation setting, one could choose industries and geographic
  locations as factors. While it is straightforward to assign {\em
    one\/} industry and {\em one\/} geographic location to a company,
  this may fail capture the dependence on further relevant industries
  and geographic locations of internationally operating firms.

  We employ {\em Bayesian variable selection (BVS)\/} methods using
  both the prior knowledge of the main industry and headquarter
  location of a firm and allowing to select further factors, while
  controlling the expected number of factors. This gives a stable
  assignment of factors to assets to be used in \eqref{eq:1}.

  To this end we impose a linear factor structure on asset returns.
  In a {\em (linear) factor model\/} (see e.g.\ Chapter 6 of
  \citep{McNeil2015}), the return vector of $p$ firms,
  $\bm r=(r_1, \ldots, r_p)^T$, is represented as
  \begin{equation*}
    r_i=\alpha_i + \beta_{i1} x_1 + \beta_{i2} x_2 + \cdots + \beta_{id}
    x_d + \varepsilon_i, \qquad i=1,\ldots, p,
  \end{equation*}
  where $x_1, \ldots, x_d$ are the {\em common factors},
  $\beta_{i1}, \beta_{i2}, \ldots, \beta_{id}$ are the {\em factor
    loadings} and
  $\bm \varepsilon=(\varepsilon_1, \ldots, \varepsilon_p)^T$ are the
  {\em idiosyncratic error terms}, assumed to be uncorrelated and with
  mean $0$. Contrary to an OLS estimator, which typically assigns
  non-zero factor loadings to all factors, methods such as {\em
    Lasso\/} (e.g.\ \citet{Hastie2009}) select a (small) subset of
  factors by assigning both zero and non-zero factor loadings. If
  identical priors for all factors are used, then Lasso could be
  employed instead of a BVS method (see \citep[Section
  4.4.2]{Fahrmeir2013} for the connection of Lasso and BVS). 
  However, given that prior knowledge about some factors
  (e.g.\ the company VW is an automotive company headquartered in
  Germany) is available, a BVS method with non-identical priors is
  preferred.

  Below we outline Bayesian model selection, the method used in this
  article. A further popular BVS method, BVS with spike and slab priors,
  is developed in \citep{George1997}. 

  For
  every firm $i$, we estimate the {\em posterior inclusion probability
    (PIP)\/} of each factor $k$ and set $\1_{k,i}=1$ if the PIP is
  greater than $1/2$. This is the so-called {\em median probability
    model\/}. \citet{Barbieri2004} show that the
median probability model is often optimal in terms of prediction. 
In our example, prior is initially set to force inclusion of a firm's
headquarter's country and its primary industry%
, all other prior
inclusion probabilities are set to include six factors on average. As
the PIP's are recalibrated periodically, the current PIP's are chosen
as the new prior inclusion probabilities. This provides a 
greater stability of the parameters over time.

  \subsection{Bayesian linear model}
\label{sec:bayes-line-model}

We consider the linear model
\begin{equation*}
  \bm y = \bm X\bm \beta + \bm \varepsilon. 
\end{equation*}
In the Bayesian setting -- see Section 3.5 \citep{Fahrmeir2013} -- we
assume that  
\begin{equation*}
  \bm y|\bm \beta, \sigma^2\sim \Ncdf(\bm X\bm \beta,
  \sigma^2\bm I),
\end{equation*}
with $\bm\beta$ and $\sigma^2$ stochastic.

A conjugate prior, i.e., where the prior and the posterior
distributions are from the same distribution family, is 
\begin{align*}
  \bm \beta|\sigma^2&\sim \Ncdf(\bm m, \sigma^2 \bm M)\\
  \sigma^2 &\sim \text{IG}(a,b),
\end{align*}
where $\text{IG}(a,b)$ denotes the inverse gamma distribution with
parameters $a, b$.
An equivalent formulation is that the pair $(\bm\beta, \sigma^2)$
follows an normal inverse gamma distribution 
\begin{equation*}
  (\bm \beta, \sigma^2)\sim \text{N-}\Gamma^{-1}(\bm m, \bm M, a, b). 
\end{equation*}
The {\em posterior distribution\/} is given as (see e.g.\
\cite{Fahrmeir2013}) 
\begin{align*}
  (\bm\beta, \sigma^2)|\bm y&\sim \text{N-}\Gamma^{-1}(\tilde{\bm m},
                          \tilde{\bm M}, \tilde a, \tilde b),
\end{align*}
where
\begin{align*}
  \tilde {\bm M} &= \left(\bm X'\bm X + \bm M^{-1}\right)^{-1}\\
  \tilde {\bm m} &= \tilde {\bm M} \left(\bm M^{-1} \bm m +
                      \bm X'\bm y\right)\\
  \tilde a &= a + \frac{n}{2}\\
  \tilde b &= b + \frac{1}{2}\left(\bm y'\bm y + \bm m'\bm
             M^{-1}\bm m - \tilde{\bm m}'\tilde{\bm
             M}^{-1} \tilde{\bm m}\right). 
\end{align*}

\subsection{Bayesian model comparison and selection}
\label{sec:model-selection}

This method for variable selection considers candidate models
$M_i$, $i=1,\ldots, m$. In the linear setting, each model $M_i$
includes a specific set of independent variables and excludes the
other variables. For the posterior model probability we have
\begin{equation*}
  p(M_i|\bm y)\propto p(\bm y|M_i) p(M_i),
\end{equation*}
with $p(\bm y|M_i)$ the so-called {\em marginal likelihood}. Define
indicator variables $\gamma_1, \ldots, \gamma_d$, with
$\gamma_k=\1_{\{\beta_k\not=0\}}$,i.e., if $\gamma_k=1$, then the
$x_k$ is included in the model. A {\em prior model \/} is then
\begin{equation*}
  p(M_i) = \prod_{k=1}^d w_k^{\gamma_k} (1-w_k)^{1-\gamma_k},
\end{equation*}
where $w_k\in [0,1]$, $k=1,\ldots, d$. If $w_k=1/2$, then
$p(M_i|\bm y)\propto p(\bm y|M_i)$.

In our setting we set $w_k=1$ initially for the industry and
location that is hard-coded for a firm (this data is available on
Bloomberg). All other 
weights $w_k$ are set to $\theta=\E[S]/d$, where $S$ is the (unknown)
model size. Under the assumption that $S$ follows a binomial
distribution $B(d,\theta)$, $\E(S) = \theta\cdot d$, so $\theta$ is
chosen to attain a target expected model size.

The marginal likelihood can be calculated from the poster density for
model $M_i$,
\begin{equation*}
  p(\bm \beta, \sigma^2|\bm y,M_i) = \frac{p(\bm y|\bm\beta,
    \sigma^2,M_i) p(\bm\beta, \sigma^2|M_i)} {p(\bm y|M_i)}, 
\end{equation*}
by re-arranging to
\begin{equation*}
p(\bm y|M_i) = \frac{p(\bm y|\bm\beta,
    \sigma^2,M_i) p(\bm\beta, \sigma^2|M_i)} { p(\bm \beta,
    \sigma^2|\bm y,M_i)}. 
\end{equation*}
This can be calculated analytically or by Markov Chain Monte Carlo
methods (MCMC) if the number of models, $2^d$, is large. The posterior
probability of $\gamma_k$ across all models is given by the {\em
  posterior inclusion probability (PIP)},
\begin{equation}
  \label{eq:3}
  \p(\gamma_k=1|\bm y) = \sum_{\beta_k\in M_i, i=1,\ldots, 2^d}
  \p(M_i|\bm y). 
\end{equation}
If MCMC is used (as in our case), then PIP's are estimated as
the frequency of visited models including the covariate relative to
the total number of visited models.

\section{Application to stock market data}
\label{sec:stress-test-scen}

In this section, we use historical data to demonstrate the
  outcome of correlation stress testing on a typical stock
  portfolio. After introducing the data, we 
first look at the factor selection and resulting parameterisation, and
then at the stress testing results.


\subsection{Data}

The methods developed in this article apply to any portfolio of risky
assets. As a showcase, we download daily equity data from Refinitiv
Eikon for the period of 1999 to 2021. The data set includes 505 constituents from the S\&P 500 and 30
constituents from the German DAX index. An equally weighted portfolio
of these assets will build the baseline. 
 
Given a portfolio of risky assets, it is necessary to select a
universe of relevant correlation risk factors, which will be the basis
for the Bayesian factor selection. We choose to stress country and
industry factors and, hence, download historical equity index data
from Refinitiv Eikon that represents these factors
(see Table~\ref{tab:indices}).\footnote{One could easily extend this
  by adding additional indices, e.g. MSCI's Small Cap, Large Cap,
  Growth, Value or Momentum indices.} More specifically, from the MSCI
All Country World Index family (ACWI) we download 8 regional indices,
including 4 mature market (MM) and 4 emerging market (EM)
indices. Industry factors are represented by 11 MSCI Global Industry
Classification Standard (GICS) sector indices.


\begin{table}
  \caption{MSCI indices that are used as correlation risk factors.
  }\label{tab:indices}
  \begin{adjustbox}{max width=\textwidth}
    \begin{tabular}{lll}
\toprule
            RIC &         Name &                                        Description \\
\midrule
 .dMINA00000PUS &  MM-Americas &                          North America price index \\
 .dMIEU00000PUS &    MM-Europe &                                 Europe price ondex \\
 .dMIPC00000PUS &   MM-Pacific &                                Pacific price index \\
 .dMILA00000PUS &  EM-Americas &         Emerging markets Latin America price index \\
 .dMIEE00000PUS &      EM-EMEA &                  Emerging markets EMEA price index \\
 .dMIMS00000PUS &      EM-Asia &                  Emerging markets Asia price index \\
 .dMIWD0EN00PUS &       Energy &                     ACWI energy sector price index \\
 .dMIWD0MT00PUS &    Materials &                  ACWI materials sector price index \\
 .dMIWD0IN00PUS &  Industrials &                ACWI industrials sector price index \\
 .dMIWD0CD00PUS &    ConsDiscr &  ACWI consumer discretionary sector price index... \\
 .dMIWD0CS00PUS &  ConsStaples &    ACWI consumer staples sector price index        \\
 .dMIWD0HC00PUS &   Healthcare &           ACWI health care sector price index      \\
 .dMIWD0FN00PUS &   Financials &                 ACWI financials sector price index \\
 .dMIWD0IT00NUS &     InfoTech &     ACWI information technology sector price index \\
 .dMIWD0TC00PUS &         Comm &  ACWI communications services sector price inde... \\
 .dMIWD0UT00PUS &    Utilities &                  ACWI utilities sector price index \\
 .dMIWD0RE00PUS &   RealEstate &                ACWI real estate sector price index \\
\bottomrule
\end{tabular}
  \end{adjustbox}
\end{table}

\subsection{Factor selection and fit}

Using the Bayesian factor selection procedure from
Section~\ref{sec:fact-select-stock}, correlation risk factors are
assigned to each asset in the portfolio. Knowledge of a
  company's primary country and industry is used by setting their
  initial prior probabilities to one, which ensures they are included
  as risk factors. These are the
country where the company is headquartered and its primary industry as
provided by Refinitiv Eikon.

Factors are re-selected at a quarterly frequency. For every firm $i$ we
estimate the {\em posterior inclusion probability (PIP)\/} of each
factor $k$ and set $\1_{k,i}=1$ if the PIP is greater than $1/2$ (cf.\
Section~\ref{sec:model-selection}). Every quarter, the previous
parameter inclusion probabilities enter as prior
probabilities. This modelling choice supports a robust
allocation of factors, yet leaves enough flexibility to add or remove
factors if the evidence at the time of selection is strong
enough. This caters to the fact that most companies have a relatively
rigid business model, but  evolve through time
and tend to occasionally enter or exit different markets and sectors.

\begin{figure}[t]
  \centering
  \includegraphics[width=.49\textwidth]{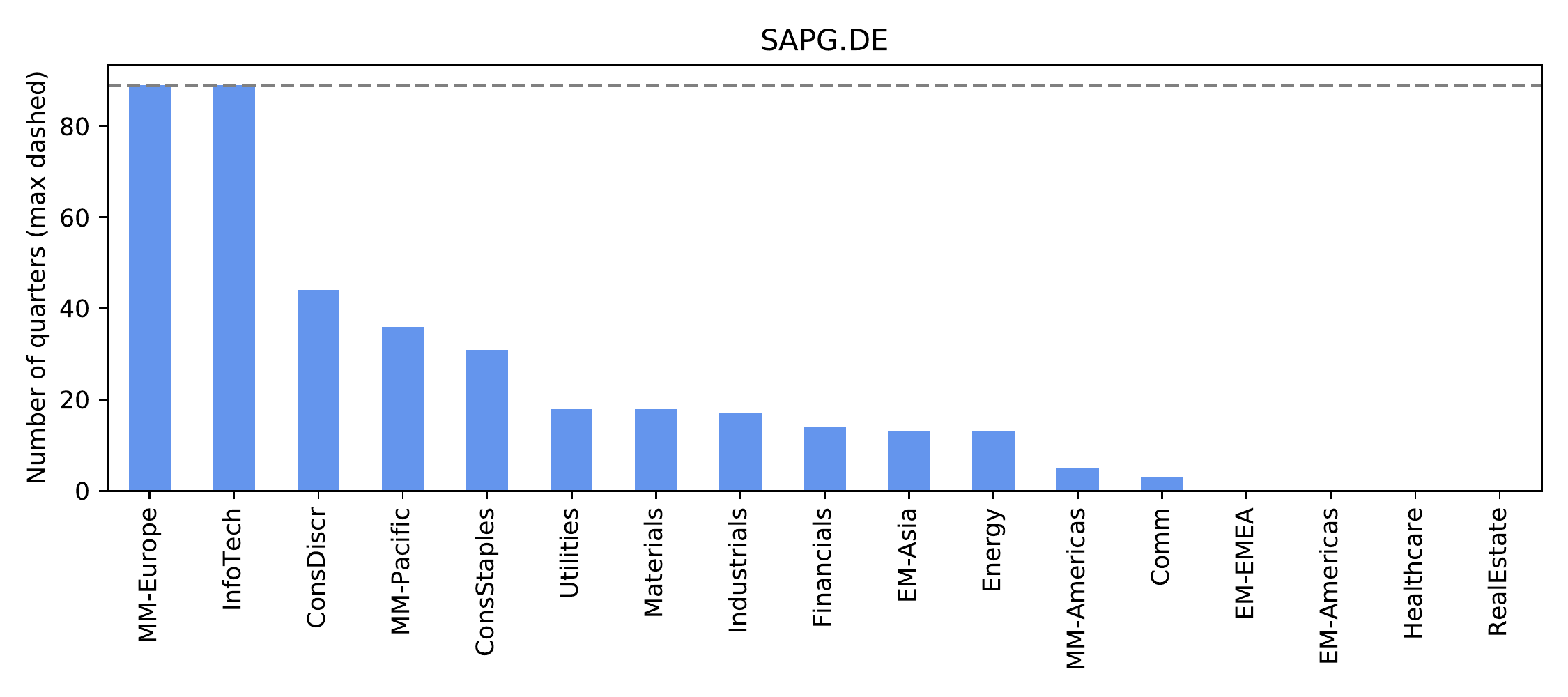}
  \includegraphics[width=.49\textwidth]{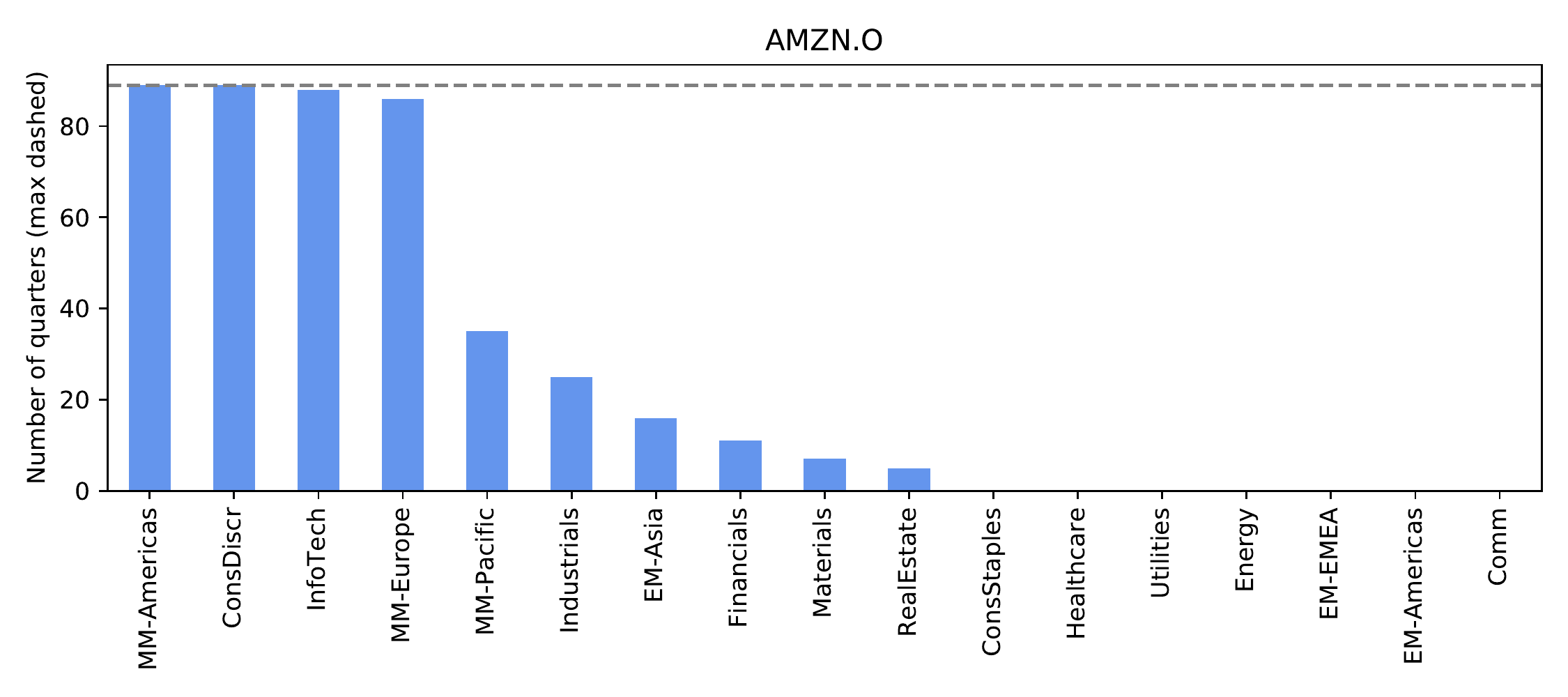}
  \includegraphics[width=.49\textwidth]{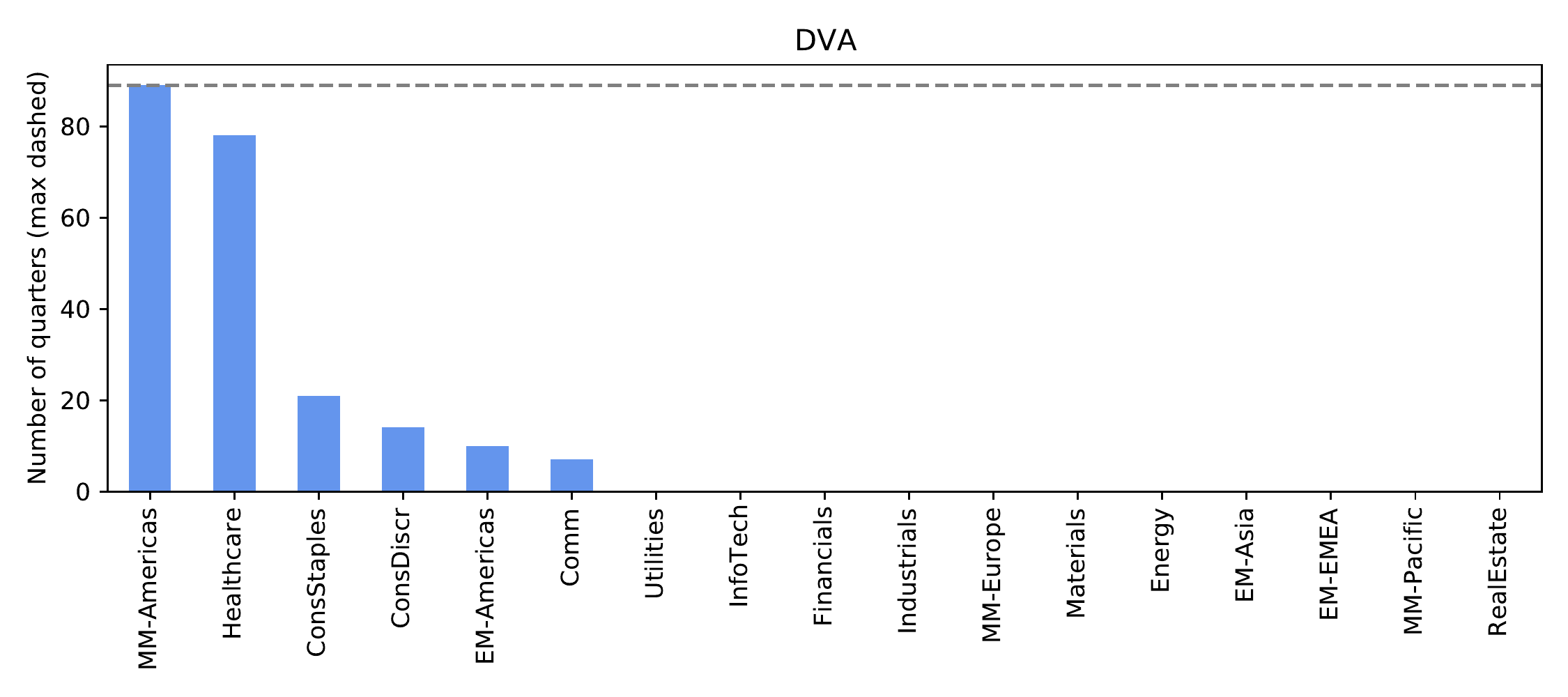}
  \includegraphics[width=.49\textwidth]{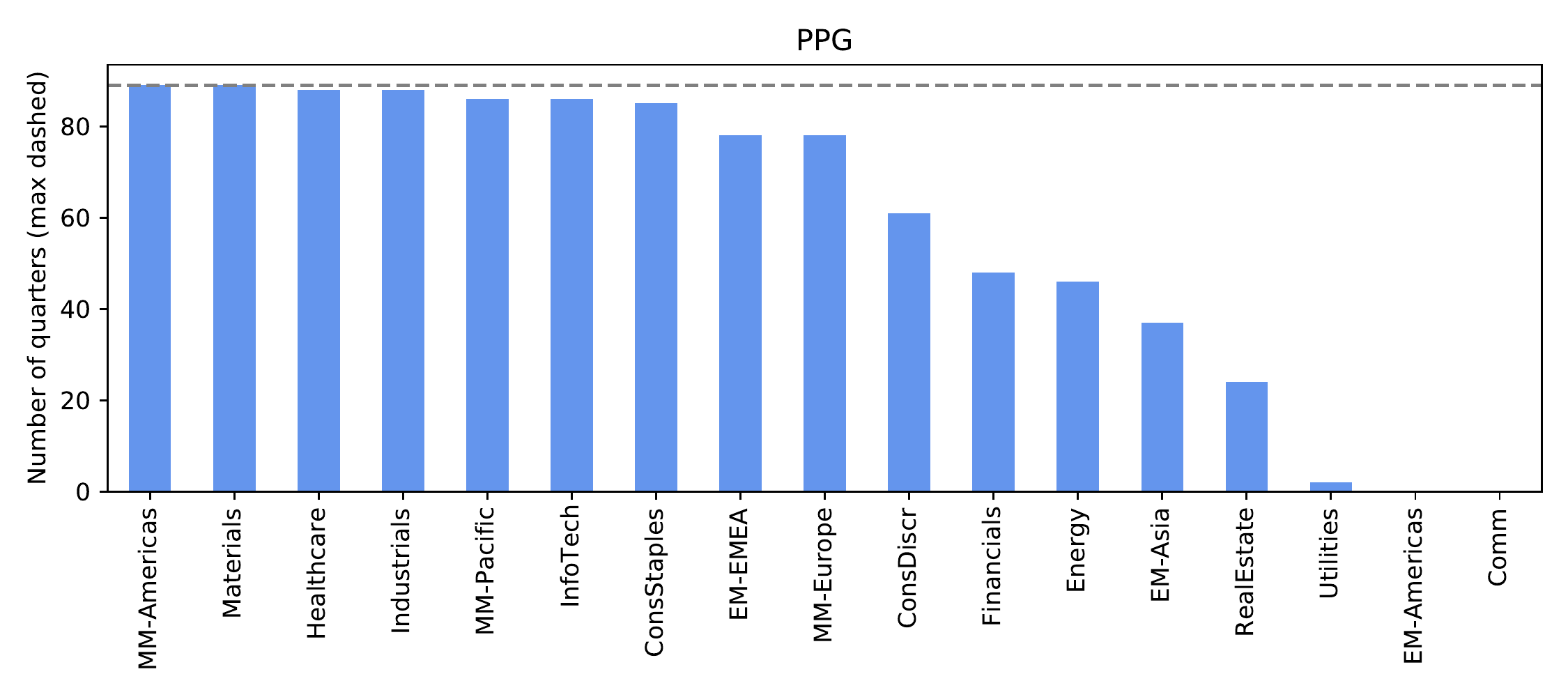}
  \caption{Correlation factor allocation. Factors are re-calibrated on
    a quarterly basis; the plots show how often a factor has been
    included. Top left: SAP is a German IT company. Top right: Amazon
    is a US based online retailer; however, it is also the world's
    largest provider of computing services (AWS) with a strong
    presence in Europe. Bottom: lowest (left) and highest (right)
    factor allocation.}
  \label{fig:factor_selection} 
\end{figure}

Figure~\ref{fig:factor_selection} shows the factor allocation for four
exemplary assets. As factors are re-calibrated on a quarterly basis,
the plots show how often a factor has been included. Of the 88
quarters in the sample, SAP SE (SAPG.DE) -- a German IT company -- has
both MM-Europe and InfoTech always included. Both factors are also the
initial prior. In contrast, the BVS consistently selected InfoTech and
MM-Europe as additional correlation risk factors for Amazon
Inc. (AMZN.O). This is a very reasonable result as Amazon is not only
a large US online retailer, but also the world's largest provider of
computing services (AWS) with a very strong presence in
Europe. Looking at the extremes, PPG Industries Inc.\ (PPG) provides
materials to a broad range of companies worldwide, thereby exposing
itself to the highest number of correlation risk factors in our
sample.

The parameters $\eta$, $\lambda_1, ..., \lambda_d$ and
$\nu_1 ..., \nu_d$ are easily determined by standard regression
techniques such as OLS on the transformed correlations
$\text{tanh}^{-1}(c_{ij})$. The fit is computationally efficient and
hence, allows the processing of very large and complex
portfolios. With parameters calibrated on a regular basis, the
parameter history can be used to better understand correlation
dynamics and to put a plausibility constraint on correlation
scenarios. Here, parameters are calibrated daily from the $250$
log-returns preceding day $t$. 
As outlined in
Section~\ref{sec:factor-model}, we test for positive-semidefiniteness
and, if necessary, use the search algorithm in
\citet{higham2002computing} to find the nearest correlation matrix
that tests positive definite.

\begin{figure}[t]
  \centering
  \includegraphics[width=\textwidth]{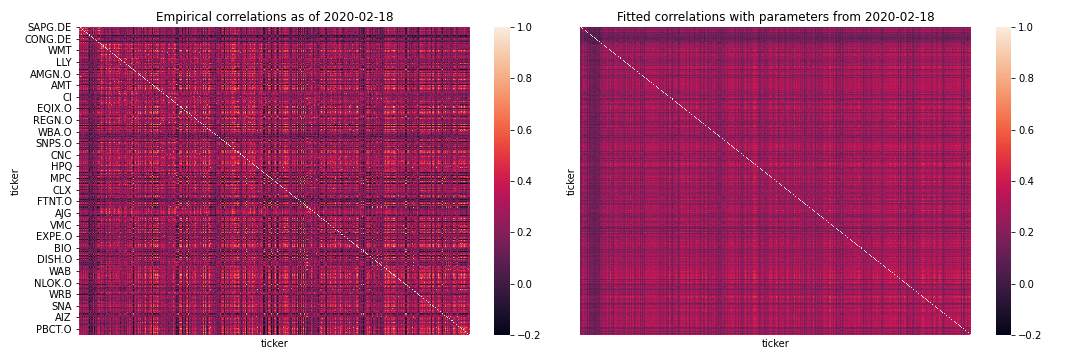}
  \includegraphics[width=\textwidth]{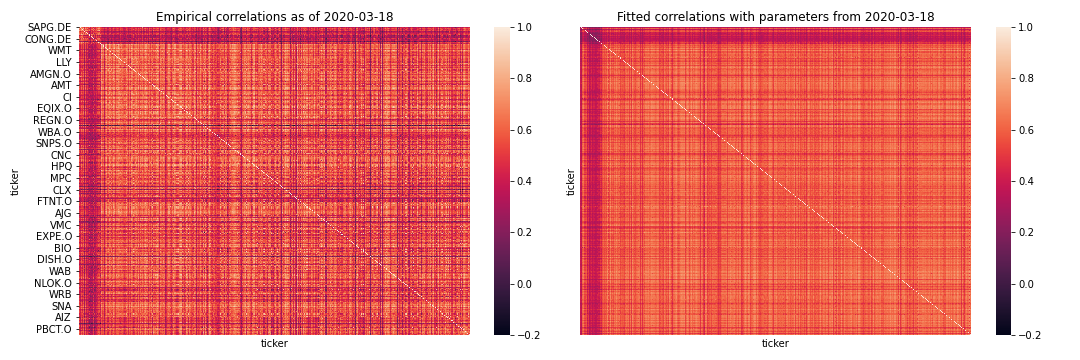}
  \caption{Heatmaps of empirical (left) and fitted (right) correlations. The spike in correlations between 18 Feb (top) and 18 Mar 2020 (bottom) was caused by markets reacting to the COVID-19 crisis. Empirical correlations are calculated on a window of 250 trading days.}
  \label{fig:heatmap_params} 
\end{figure}

The heatmaps in Figure~\ref{fig:heatmap_params} show empirical correlations over 250 trading days (left)
as well as the corresponding fitted correlation matrices (right),
where a brighter colour indicates a higher correlation. The model is capable of
capturing a number of correlation structures that are visible as
shaded areas or stripes in all heatmaps. The top rows and left most
columns show correlations between German DAX assets and US S\&P 500
assets. Naturally, cross-country correlations are structurally lower
than within country correlations.

Owing to the COVID-19 outbreak, New York City -- one of the world's
largest financial centres -- started to lock down on Friday, 13
  March 2020. The following days saw
some of the largest market drops in history. As is symptomatic for
falling markets, correlations spiked during the downturn. The
  heatmaps, being depicted on an identical color scale,
show this by jumping from a dark purple (top) to a bright orange
(bottom). We will later show that this constituted, at least
partially, the worst-case scenario of the portfolio. In other words, diversification
benefits diminished at times where they would have been needed most.

\begin{figure}[t]
  \centering
  \includegraphics[width=\textwidth]{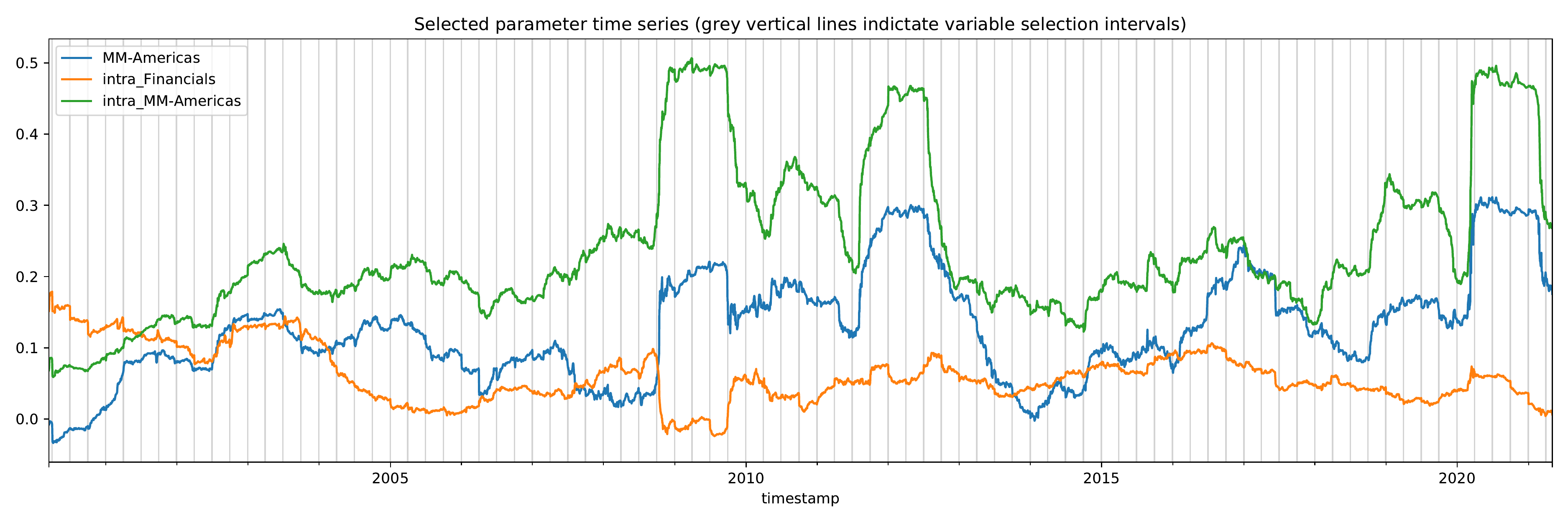}
    \includegraphics[width=\textwidth]{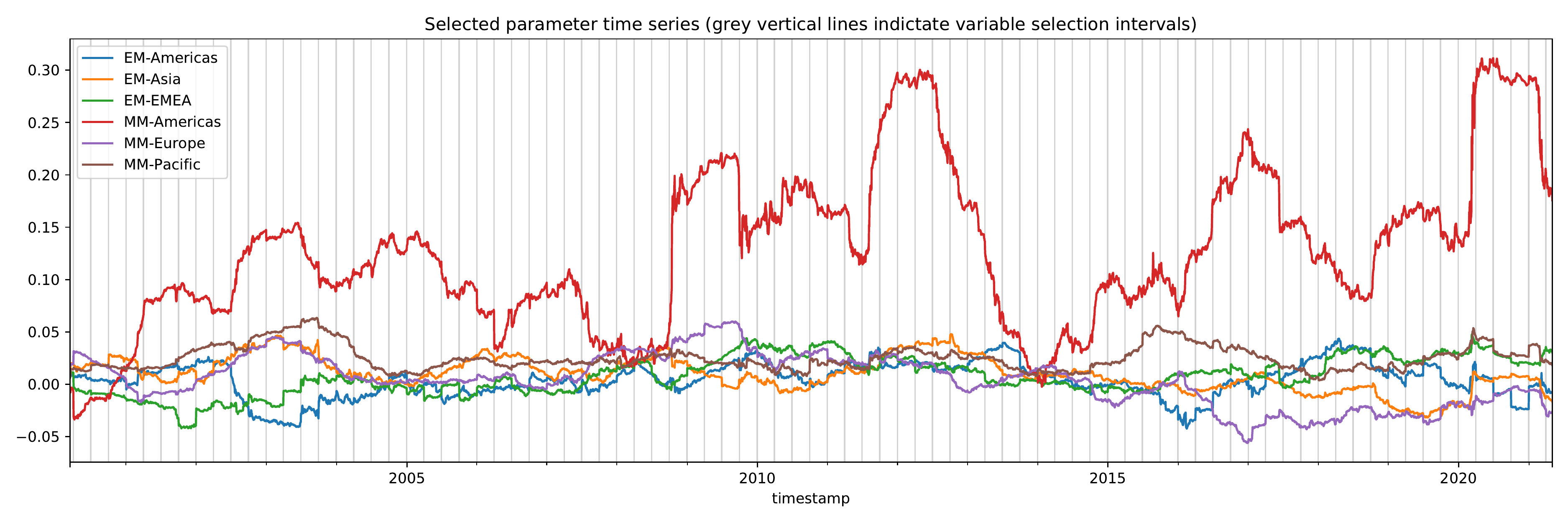}
  \includegraphics[width=\textwidth]{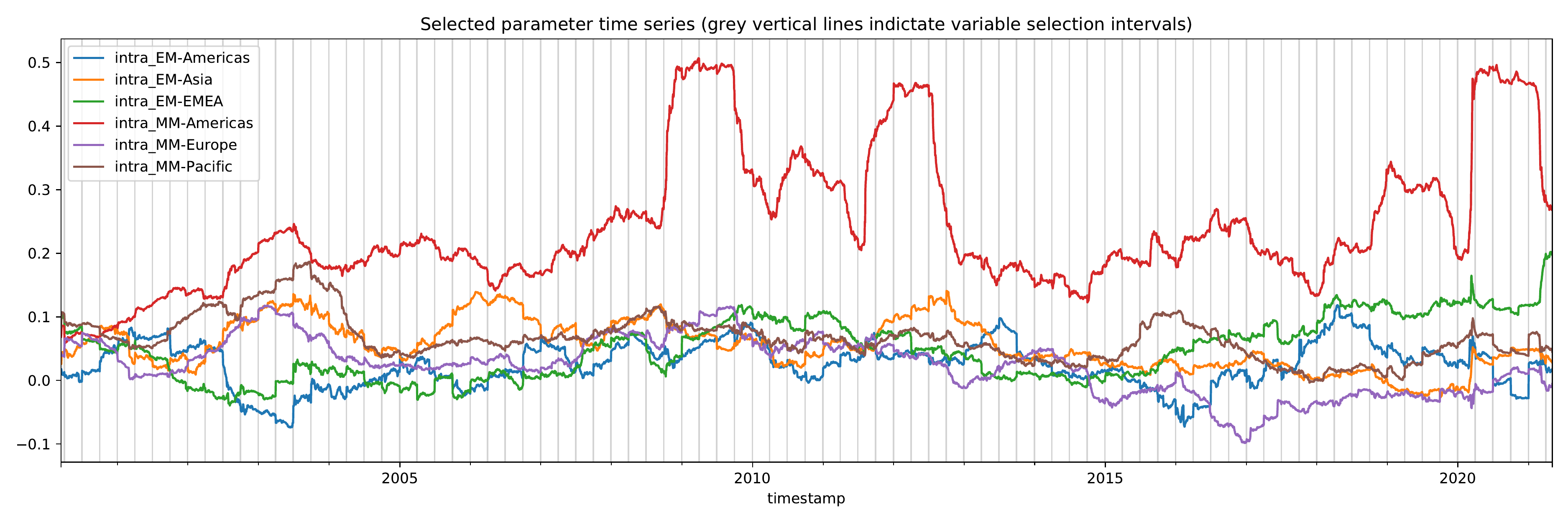}
  \caption{Top: fitted parameters for risk factors with high
    loads. Bottom left: fitted ``inter'' parameters for regional risk
    factors (``$\lambda_k$''). Bottom right: fitted ``intra''
    parameters for regional risk factors (``$\nu_k$''). Grey vertical
    lines indicate factor selection intervals.}
  \label{fig:ts_params} 
\end{figure}

The time series of fitted correlation parameters in
Figure~\ref{fig:ts_params} shows correlation dynamics over time. One
can clearly see the spikes in correlation during the financial crisis,
the government debt crisis and most recently the COVID-19
pandemic. The correlation dynamics within the financial sector during
2008 are particularly interesting. The factor load on financials is
almost entirely consumed by global factors as soon as the crisis
spills over into the whole economy. To this end, recall the
summation formula in~\eqref{eq:tanhsum}, which approximates
correlations in our model by sums of individual correlation risk
factor loads. It is therefore 
  natural that the (financial) sector specific correlation structure
  was temporarily subsumed by the more common Americas factors.
For the 250 trading days leading up to the default of Lehman
  Brothers, the average correlation among all assets was 0.32 while
  the correlation within the financial sector was 0.40. The model
  would therefore add sector specific correlation through the
  corresponding ``intra\_Financials'' factor load, which at the time was
  roughly 0.1. During the following 250 trading days, i.e.\ during the
  financial crisis, the average correlation increased to a level of
  0.50 overall and 0.51 for financials, which explains the increase in
  ``Americas'' and decrease in ``Financials'' factor loads.

Figure~\ref{fig:boxplots_stress} presents box-plots of coefficients of
correlations between factors (left) and within factors
(right). Unsurprisingly, the factor representing correlations within
North America (``intra\_MM-Americas'') captures the majority of the
correlation dynamics in our portfolio, followed by the factor
capturing correlations between North America and other countries. This
shows that the method adapts well to the underlying portfolio, which
comprises only US and German assets.

In general, intra-correlations are higher than
inter-correlations. This is not surprising as correlations are higher
for similar assets, such as assets within the same country or industry.

\subsection{Stress test results}


\begin{figure}[t]
  \centering
  \includegraphics[width=\textwidth]{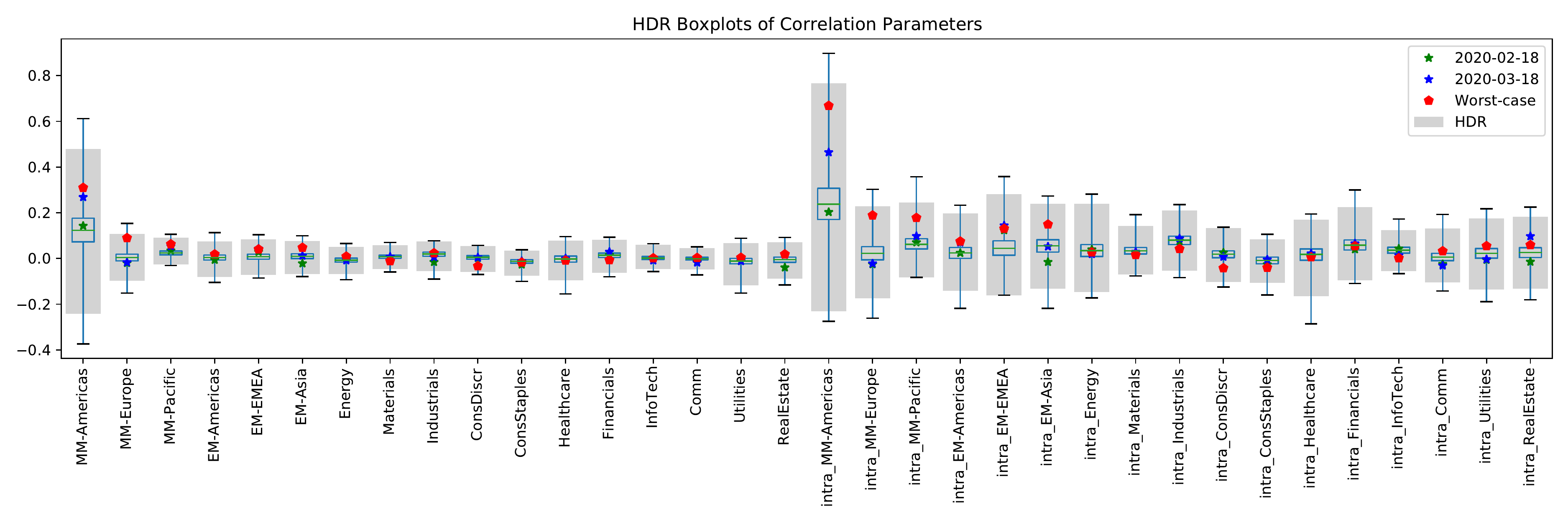}
  \caption{Reverse stress test, i.e., worst-case scenario within the
    $95\%$ highest density region as of 2020-02-18. Triangles indicate
    stress scenarios from Monte Carlo and historical simulation. The
    stress scenario was partially realized by 2020-03-18. Stars
    indicate correlation parameters for specific dates. The large
    shift in correlations around that time is owed to the COVID-19
    crisis.}
  \label{fig:boxplots_stress} 
\end{figure}

\begin{figure}[t]
  \centering
  \includegraphics[width=\textwidth]{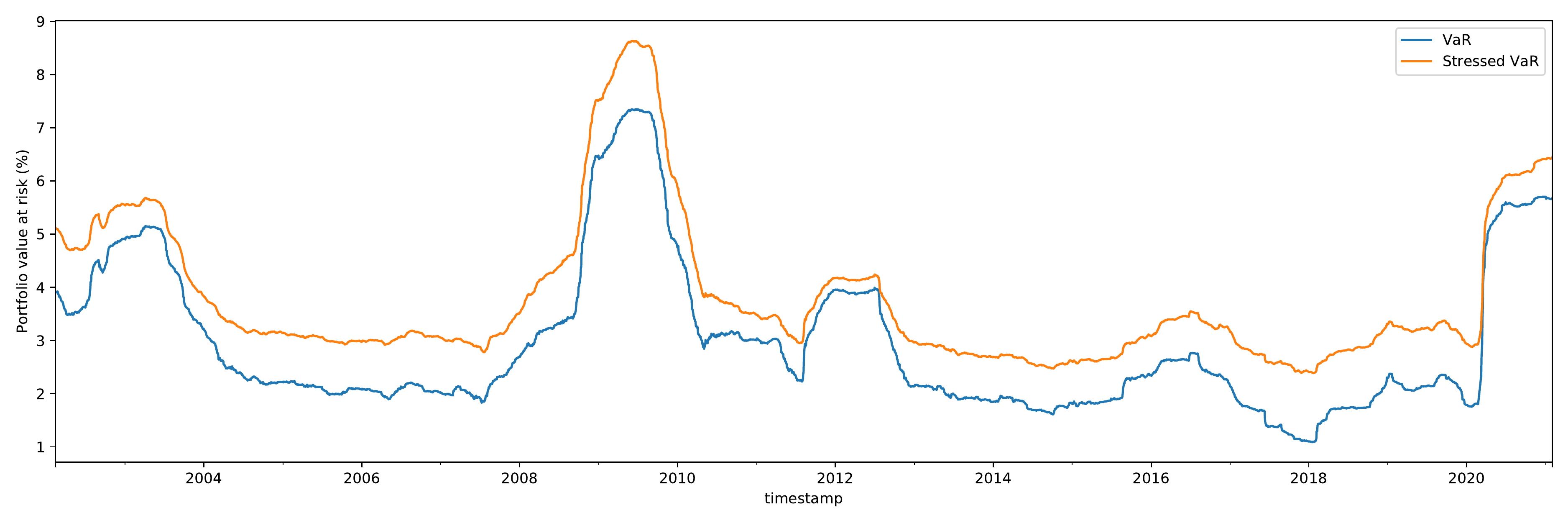}
  \caption{One day 99\% VaR and stressed VaR. The stressed war uses
    the reverse stress scenario of 2021-05-04. The difference between
    both VaRs diminishes occasionally, indicating that the stress
    scenario has been realized.}
  \label{fig:var_timeseries} 
\end{figure}

A shift in correlation has, ceteris paribus, no {\em instantaneous\/} effect on a
portfolio's value, therefore, to reveal the impact of a correlation
stress test requires calculating portfolio risk measures.
Value-at-risk (VaR) in a {\em variance-covariance approach} has been
proposed in Section~\ref{sec:corr-stress-test} as a straightforward
portfolio risk measure.

In order to impose a plausibility constraint on the correlation stress
scenario we fit the correlation parameters to a multivariate
normal-inverse Gaussian (NIG) distribution (cf.\
Section~\ref{sec:reverse-stress-test}). The (multivariate) NIG
distribution belongs to the family of normal-mean-variance mixtures,
which generalise the (multivariate) normal distribution. It allows for
skewness in the margins as well as a higher variation in tail
behaviour compared to the normal distribution, while still being a
light-tailed distribution, which is appropriate for the correlation
parameters.  For reasons of computation time, the NIG distribution is
calibrated using every 10\textsuperscript{th} observation from the
parameter data set, ie., 552 samples over time. Calibration is done
using the expectation-maximization (EM) algorithm described in
\citet[Chapter 3]{McNeil2005}, which goes back to
\citet{Dempster1977}. We use a Kolmogorov–Smirnov test to assess the
quality of the calibration. The null hypothesis of the test is not
rejected for 19 out of 34 marginal distributions at the 5\% level,
despite being fit to 34-dimensional
data. 
The boxplots in Figure~\ref{fig:boxplots_stress} show the range
(whiskers) and inter quartile range (box) of the correlation
parameters from the NIG distribution. The highest density region that
is relevant for the stress test is represented by a grey area.

Figure~\ref{fig:boxplots_stress} shows the worst correlation stress
scenario at the 95\% level. Two approaches are used to determine the
stress scenario. First, historical simulation, where each empirically
observed set of correlation parameters is associated with a portfolio
risk metric (here VaR). The parameter constellation that yields the
highest risk within the 95\% quantile is indicated by a right pointing
triangle. Second, Monte Carlo simulation, where we sample from the
continuous parameter distribution. The sample space is then restricted
to all scenarios that fulfil $f(x) \geq f_{0.05}$ (cf.\
Section~\ref{sec:reverse-stress-test}). Within that restricted sample,
the parameter constellation that produces the highest risk is
indicated by a left pointing triangle.

Both, historical simulation and Monte Carlo simulation yield similar
scenarios. However, the Monte Carlo scenario is more extreme, as it
reaches a broader range of potential parameter constellations. In any
case, the worst scenario for the test portfolio is always an increase
in correlations. This result is intuitive as the portfolio at hand
only benefits from na\"ive diversification. A hedged portfolio, on the
other hand, would likely suffer under decorrelation scenarios.

Stars in Figure~\ref{fig:boxplots_stress} represent correlation
parameters for specific dates. One can see that on 2020-02-18 most
parameters were close to the center of their distribution. One month
later, on 2020-03-18, the worst case has been partially realized, as
indicated by the stars shifting closer to the triangles. The
parameter constellation of the correlation stress scenario
itself remains unchanged because it depends primarily on the portfolio
weights and the correlation risk factors associated with the portfolio
constituents. This consistency of the worst case scenario is a welcome
result, from a risk management perspective, as it shows that the
method is capable of providing stable guidance on the risks of a
specific portfolio.

Figure~\ref{fig:var_timeseries} shows the 1-day $\text{VaR}_{99\%}$
with and without stressed correlations. All VaR's are calculated
  using Equation \eqref{eq:var}. Stressed VaR uses the
correlation matrix from the $95\%$-stress
scenario of 2021-05-04, the last day in the data.\footnote{The
  scenario itself is very similar to the scenario shown in
  Figure~\ref{fig:boxplots_stress} as `Worst-case (MC)'. Naturally,
  given the large number of US assets in the data, the worst case
  scenario is dominated by increasing correlations among `Americas'
  assets.} 
As stressed VaR measures
the impact of a plausible correlation stress scenario on historical
data, the difference between stressed VaR and unstressed VaR gives
insights into the severity of this scenario in the recent
history. Loosely speaking, the difference in VaR's can be
attributed to correlation risk. The distance 
between both VaRs is highest during ``normal'' markets (i.e.,
  when VaR is comparably low), where it often
exceeds 50\% of the unstressed VaR. This is a significant value, given
that the stressed portfolio is equally weighted and, thus, only
benefits from na\"ive diversification. A portfolio that benefits
  from greater diversification, e.g.\ because it is
  ``optimally-weighted'' or hedged to minimize risks, would presumable
  react more sensitive to correlation changes.

During distress periods, the distance between VaR and stressed VaR
diminishes, indicating that the stress scenario is at least partially
realized. This can be observed during the 2008 financial crisis, the
subsequent government debt crisis and most recently during the 2020
COVID-19 crisis.



\section{Conclusion}\label{sec:conclusion}

Correlation, as the driver of
diversification, has been extensively studied in the finance
literature. For example, multivariate time series models such as
DCC-GARCH capture the time variation of correlation. The more recent
literature uses hierarchical clustering methods to infer
dependence networks from correlation data, which allows to determine
economic linkages between assets. However, other than that the
literature on linking economic risk factors and correlations is
scarce. Likewise, stress testing of correlations by translating an
economic scenario into a change of correlations, is largely an
unexplored field. 

We develop a flexible correlation stress testing framework that links
risk factors with asset correlations. The process consists of three
steps:

First, the correlation matrix of asset returns is specified as
a parametric function of risk factors, involving both ``intra''- and
``inter''-correlations amongst the risk factors. ``Intra''-correlations
are correlation contributions where assets share a risk factor, and
``inter''-correlations are correlation contributions where assets do
not share a risk factor. The functional form is calibrated to market
data in two steps: Identifying an assets' relevant risk factors
is achieved using Bayesian variable selection, which allows to specify
prior information (such as the country of the firm's headquarter) as
well as control the number of factors selected. 
The risk factor loadings are calibrated from an empirical correlation
matrix.

Second, scenarios, in particular stress scenarios, are applied by
adjusting the risk factor loadings. The impact from the correlation
stress scenario is determined as the change to a risk measure such
as value-at-risk, expected shortfall or any other measure that employs
a correlation matrix. 

Third, with correlation factor loadings calibrated on a regular basis,
the parameter history can be used to better understand correlation
dynamics and to put a plausibility constraint on correlation
scenarios. In particular, reverse stress tests can be conducted,
identifying critical risk factor scenarios for the portfolio at
hand. The idea is to fit the risk factors loadings to a probability
distribution and define plausible scenarios as those that lie within a
given confidence level. In a multivariate setting, the range of
plausible scenarios can be identified as those that lie within a
certain {\em Mahalanobis distance\/} or {\em highest 
  density regions} (HDR) of the average scenario. In this sense, the
Mahalanobis distance and HDR can be thought of as multivariate
generalisations of quantiles. The scenario with the highest risk
(value-at-risk) within the confidence level represents the outcome of 
the reverse stress test: it is both extreme and plausible. Being able
to identify the relevant risk factors of the scenario provides
valuable information for portfolio managers and risk managers in
order to understand the correlation risk drivers of their portfolio.

In an extensive empirical example, we calculate the stressed
value-at-risk over time and identify worst-case stress
scenarios. We find that these scenarios are extreme enough to pose a
relevant threat from a risk management perspective, yet are common
enough to be realized on several occasions in our data history. 

The framework developed in this paper can be employed by any
stakeholder wishing to identify the economic risk of adverse
correlation changes. The method is particularly relevant to
regulators, for example when approving banks' capital models that
contain factor models. Future directions of research may lead to
extensions involving other kinds of risk factors, such as
macroeconomic factors or latent factors (e.g.\ principal components).

\appendix

\section{Normal-inverse Gaussian (NIG) distribution}
\label{sec:norm-inverse-gauss}

The NIG distribution arises as a special case of so-called
{\em normal-mean-variance mixtures (NMVM)}, and more specifically as a  
special case of the family of {\em Generalized Hyperbolic (GH)} 
distributions. NMVM combine a number of useful properties, amongst
them their flexibility in modelling skewness and heavy tails as well
as their tractability, both for numerical calculations and simulation
purposes. We refer to Section 3.2.2 of \citep{McNeil2005} for more
details. 

\begin{definition}
The random vector $\mathbf {X}$ is said to have a (multivariate)
normal mean-variance mixture distribution if
\begin{equation*}
  \mathbf X \stackrel{\mathcal L}{=} \mathbf m(W) + \sqrt{W} A \mathbf Z,
\end{equation*}
where
\begin{enumerate}[(i)]
\item $\mathbf Z \sim \Ncdf_k (\mathbf 0, I_k)$;
\item $W\geq 0$ is a non-negative, scalar-valued random variable
  independent of $\mathbf Z$;
\item $A\in \R^{d\times k}$ is a matrix;
\item $\mathbf m:[0,\infty)\rightarrow \R^d$ is a measurable
  function. 
\end{enumerate}
\end{definition}
We have
\begin{equation*}
  \mathbf X|W=w \sim \Ncdf_d(\mathbf m(w), w\Sigma),
\end{equation*}
where $\Sigma=A A'$. A possible concrete specification of $\mathbf
m(W)$ is
\begin{equation}
  \label{eq:9}
  \mathbf m(W) = \mathbf \mu + W\mathbf \gamma,
\end{equation}
where $\mathbf\mu$ and $\mathbf\gamma$ are vectors in $\R^d$. If
$\mathbf\gamma=0$, then the distribution is a NVM. 

A special case are the {\em generalized hyperbolic (GH)
  distributions}, which are NMVM's with mean specification (\ref{eq:9})
and mixing distribution $W\sim N^{-}(\lambda, \chi, \psi)$, a
generalised inverse Gaussian (GIG) distribution. We write $\mathbf
X\sim \text{GH}_d(\lambda, \chi, \psi, \mathbf \mu, \Sigma,
\mathbf\gamma)$. The specification is not unique in the sense that
scaled versions of the parameters describe the same
distribution.

The NIG distribution arises as the special case where
$\lambda=-1/2$. An extensive treatment of the NIG distribution is
found in \citep{BarndorffNielsen1997}. Amongst other useful
properties, closed formulas for the moment-generating function exist,
so all moments are easily calculated; linear combinations of NIG
variables are again NIG-distributed; the NIG distribution features
infinite divisibility, giving rise to the NIG L\'evy process, which
may be represented as a Brownian motion with a random time change. 

\bibliographystyle{abbrvnamed} %
\bibliography{finance} %

\end{document}